\documentclass[prb,twocolumn,showpacs,reprint,preprintnumbers,amsmath,amssymb]{revtex4}

\usepackage{color}
\usepackage{graphicx}
\usepackage{dcolumn}
\usepackage{bm}

\preprint{submitted to Physical Review B}

\begin{document}

\title{First-principles calculations of exchange interactions, spin waves, and temperature dependence of
magnetization  in inverse-Heusler-based spin gapless semiconductors }

\author{A. Jakobsson$^{1,2}$}\email{adam.jakobsson@physics.uu.se}
\author{P. Mavropoulos$^2$}\email{ph.mavropoulos@fz-juelich.de}
\author{E. \c{S}a\c{s}{\i}o\u{g}lu$^2$}\email{e.sasioglu@fz-juelich.de}
\author{S. Bl\"{u}gel$^2$}
\author{M. Le\v{z}ai\'{c}$^2$}
\author{B. Sanyal$^1$}
\author{I. Galanakis$^3$}\email{galanakis@upatras.gr}

\affiliation{$^1$Department of Physics and
Astronomy, Uppsala University, Box 516, 75120 Uppsala, Sweden \\
$^2$Peter Gr\"{u}nberg Institut and Institute for
Advanced Simulation, Forschungszentrum J\"{u}lich and JARA, 52425
J\"{u}lich, Germany \\
$^3$Department of Materials Science, School of
Natural Sciences, University of Patras,  GR-26504 Patra,
Greece}

\begin{abstract}
  Employing first principles electronic structure calculations in
  conjunction with the frozen-magnon method we calculate exchange
  interactions, spin-wave dispersion, and spin-wave stiffness
  constants in inverse-Heusler-based spin gapless semiconductor (SGS)
  compounds Mn$_2$CoAl, Ti$_2$MnAl, Cr$_2$ZnSi, Ti$_2$CoSi and
  Ti$_2$VAs. We find that their magnetic behavior is similar to the
  half-metallic ferromagnetic full-Heusler alloys, i.e., the
  intersublattice exchange interactions play an essential role in the
  formation of the magnetic ground state and in determining the
  Curie  temperature, $T_\mathrm{c}$. All compounds,
  except Ti$_2$CoSi possess a ferrimagnetic ground state. Due to the
  finite energy gap in one spin channel, the exchange interactions
  decay sharply with the distance, and hence magnetism of these SGSs can be
  described considering only nearest and next-nearest neighbor
  exchange interactions. The calculated spin-wave dispersion curves
  are typical for ferrimagnets and ferromagnets.  The  spin-wave stiffness constants
  turn out to be larger than those of the elementary
  3$d$-ferromagnets.   Calculated exchange parameters are used as input
  to determine the temperature dependence of the magnetization and
  $T_\mathrm{c}$ of the SGSs. We find that the $T_\mathrm{c}$ of all
  compounds is much above the room temperature. The calculated
  magnetization curve for Mn$_2$CoAl as well as the Curie temperature
  are in very good agreement with available experimental data. The
  present study is expected to pave the way for a deeper understanding
  of the magnetic properties of the inverse-Heusler-based SGSs and
  enhance the interest in these materials for application in
  spintronic and magnetoelectronic devices.
\end{abstract}


\maketitle

\section{Introduction}
\label{sec1}

Heusler compounds and alloys are a huge family of intermetallic
compounds\cite{landolt,landolt2} and they owe their name to the
German metallurgist Friedrich Heusler who in 1904 studied the
thermodynamic properties of Cu$_2$MnAl.\cite{Heusler} The
development of the research fields of magnetoelectronics and
spintronics\cite{ReviewSpin} intensified the interest in the
half-metallic Heusler
compounds.\cite{ReviewHM,FelserRev,FelserRev2} In half-metallic
magnets a metallic majority-spin electronic band structure and a
semiconducting minority-spin electronic band structure
coexist.\cite{deGroot,Pickett07} Such compounds could lead to the
creation of a fully spin-polarized current, maximizing the
efficiency of spintronic devices.\cite{Bowen} In addition,
half-metallicity in Heusler compounds is always accompanied by the
so-called Slater-Pauling rule where the total spin magnetic moment
scales linearly with the number of valence electrons in the unit
cell.\cite{Galanakis02a,Galanakis02b,Gillessen2008,GSP,Gillessen2009,Galanakis13b}
Although a large number of half-metallic Heusler compounds
compounds has been studied, still novel properties are discovered
among Heusler alloys, e.g. ferromagnetic or ferrimagnetic
semiconducting behavior,\cite{SFM,SFM2,Galanakis11,Jia2014b}
paving the way for diverse applications in
spintronics/magnetoelectronics.\cite{Perspectives}

\begin{figure}[t]
\includegraphics[width=\columnwidth]{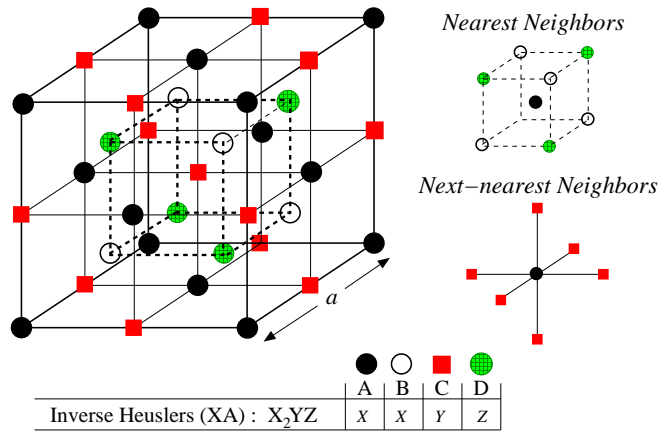}
\vskip -0.1cm \caption{Schematic representation of the lattice
structure of inverse Heusler compounds having the chemical formula
X$_2$YZ where X and Y are transition metal atoms (with the valence
of Y larger than of X) and Z is an sp-element. On the right we
present the nearest and next-nearest neighbors of an A site. Note
that the large cube contains exactly four primitive unit
cells.}\label{Structure}
\end{figure}

A class of materials bridging the gap between half-metals and
magnetic semiconductors are the so-called spin-gapless
semiconductors (SGS); magnetic semiconductors where there is an
almost zero-width energy gap at the Fermi level in the
majority-spin direction and a usual energy gap in the other
spin-direction.\cite{GS} In SGS (i) the mobility of carriers is
considerably larger than in usual semiconductors, (ii) excited
carriers include both electrons and holes which can be 100\%
spin-polarized simultaneously, and (iii) a vanishing amount of
energy is enough to excite majority-spin electrons from the
valence to the conduction band. Several compounds have been
identified as
SGS,\cite{Wang,Wang2,Kim2014,Pan10,Pan11,Liu12,Wang13,He13,VAs}
and  among them exist a few Heusler
compounds,\cite{Galanakis13b,GalanakisSGS,Xu13,Gao13,Bainsla}
 and it is
Mn$_2$CoAl, an inverse full-Heusler compound, which has attracted
most of the attention due to its successful growth in the form of
films. First in 2008 Liu and collaborators synthesized using an
arc-melting  technique Mn$_2$CoAl and found that it adopted the
lattice structure of inverse full-Heuslers (see Fig.
\ref{Structure} for a schematic representation of the structure)
with a lattice constant of  5.8388 \AA\ and a total spin magnetic
moment of 1.95 $\mu_\mathrm{B}$ per formula unit.\cite{Liu08} The
lattice of inverse Heuslers is known as the XA(or
$X_\alpha$)-structure and it is  similar to the well-known
$L2_\textrm{1}$ structure of full Heusler compounds like
Co$_2$MnAl where only the sequence of the atoms in the unit cell
changes. Moreover electronic structure calculations yielded a
ferrimagnetic state with a total spin magnetic moment of 1.95
$\mu_\mathrm{B}$ per formula unit and an antiparallel coupling
between the Mn nearest-neighboring atoms.\cite{Liu08} Although the
calculated structure in Ref. \onlinecite{Liu08} is that of a SGS,
authors do not mention it in their article.\cite{Liu08} In 2011
Meinert and collaborators studied again theoretically this
compound and almost reproduced the calculated results of Liu
\textit{et al.} using a different electronic structure
method.\cite{Meinert11} They have also calculated the exchange
constants showing that they are short range and the magnetic state
is stabilized mainly due to the direct interaction between
nearest-neighbors and predicted a Curie temperature of 890
K.\cite{Meinert11} But it was not until 2013, when Ouardi et al
identified the SGS behavior of Mn$_2$CoAl and have confirmed it
experimentally in bulk-like pollycrystalline films.\cite{Ouardi}
The experimental lattice constant was found to be 5.798 \AA , the
Curie temperature was measured to be about 720 K and the total
spin magnetic moment per formula unit was found 2 $\mu_\mathrm{B}$
at a temperature of 5 K.\cite{Ouardi} Following this research
work, Jamer and collaborators have grown thin films of 70nm
thickness on top of GaAs,\cite{Jamer13} but these films were found
to deviate from the SGS behavior.\cite{Jamer14} On the contrary,
films -grown on top of a thermally oxidized Si substrate- were
found to be SGS with a Curie temperature of 550 K.\cite{Xu2014}
First-principles calculations of Skaftouros et al identified among
the inverse Heusler compounds four additional potential SGS
materials: Ti$_2$CoSi, Ti$_2$MnAl, Ti$_2$VAs and Cr$_2$ZnSi, the
latter three being also fully-compensated ferrimagnets, and
V$_3$Al for which one V sublattice is not magnetic and the other
two form a conventional antiferromagnet.\cite{GalanakisSGS} The
SGS character of Ti$_2$MnAl was also confirmed by Jia \textit{et
al.}\cite{Jia2014} Wollman \textit{et al.}\cite{Wollmann}
confirmed the conclusion of Meinert \textit{et al.} that direct
exchange interactions are responsible for the magnetic order in
Mn$_2$CoAl studying a wide range of Mn$_2$-based Heusler compounds
and predicted a Curie temperature of 740 K using the spherical
approximation.\cite{SPA} Skaftouros \textit{et al.} have discussed
in detail the behavior of the total magnetic moment in inverse
Heusler compounds including the SGS materials.\cite{GSP} Galanakis
and collaborators have shown that defects keep the half-metallic
character of Mn$_2$CoAl but destroy the SGS
character.\cite{Galanakis2014} Finally, recent studies on the
effect of doping of Mn$_2$CoAl with Co, Cu, V and
Ti,\cite{Zhang13} as well as the anomalous Hall effect have
appeared in literature.\cite{Kudrnovsky}

Motivated by these recent advances in SGS systems, the present
work aims at a prediction and understanding of their magnetic
properties at elevated temperatures. We employ density-functional
theory at the ground state augmented by a Heisenberg model
Hamiltonian for the prediction of the temperature dependent
magnetization.

We present calculations of exchange interactions, spin waves and
temperature dependence of the magnetization in five inverse
Heusler compounds known to present spin-gapless semiconducting
behavior studied in Ref. \onlinecite{GalanakisSGS}: Mn$_2$CoAl,
Ti$_2$CoSi, Ti$_2$MnAl, Ti$_2$VAs and Cr$_2$ZnSi. We find that
magnetic behavior of the  SGSs  is similar to the half-metallic
ferromagnetic full-Heusler alloys, i.e., the intersublattice
exchange interactions play an essential role in the formation of
the magnetic ground state and in determining the critical
temperature, $T_\mathrm{c}$. Note, that even in the case of zero
total spin magnetic moment in the unit cell, the compounds under
study are fully-compensated ferrimagnets and not conventional
antiferromagnets, and thus the critical temperature should be
called ``Curie temperature'' and not ``N\'eel temperature''.  It
turns out that the $T_\mathrm{c}$ of all compounds is much above
the room temperature. The calculated magnetization curve for
Mn$_2$CoAl as well as the Curie temperature are in very good
agreement with available experimental data. The rest of the paper
is organized as follows: In Sec.\,\ref{sec2} we present the
computational method.  In Sec.\,\ref{sec3} we present the
computational results,  and Sec.\,\ref{sec3} gives the
conclusions.

\section{Computational method}
\label{sec2}

\subsection{Crystal structure and ground state calculations}
\label{sec2_2}

Prior to discussing the structure of the compounds under study, we
should note that Mn$_2$CoAl has been already synthesized
experimentally,\cite{Ouardi} and for the other four compounds one
should calculate the formation enthalpy to establish their
possible experimental existence. In Refs.
\onlinecite{Gillessen2008} and \onlinecite{Gillessen2009}, authors
studied the formation enthalpies for 810 Heusler compounds and
concluded that the formation of the compounds is not favored with
respect to the constituent elements only in the case where the
heavier main-group metals such as thallium, lead and bismuth are
involved. Thus we expect the compounds under study to be
thermodynamically stable.

All five compounds under study are called inverse Heuslers and
crystallize in the so-called XA or X$_\alpha$ structure, the
prototype of which is CuHg$_2$Ti, but usually the sequence of the
atoms follows the chemical formula X$_2$YZ. X and Y are transition
metal atoms with the valence of X being smaller than Y, and Z is a
sp atom. A schematic representation of the structure is given in
Fig. \ref{Structure}. There are four atoms along the diagonal of
the cube following the sequence X-X-Y-Z and thus the two X atoms
sit at sites of different symmetry. We use the superscripts A and
B to distinguish them. As shown also in the figure, each atom at
the A or C site resides at the center of a cube with nearest
neighbors four B sites and four D sites sitting at the corners of
the cube; equivalently each atom at a B or D site has four atoms
at A sites and four atoms at C sites as nearest neighbors.
Concerning the next-nearest neighbors each A site has as second
neighbors six C sites (and each C site has six A sites as second
neighbors), and equivalent is the situation for the B and D sites.
The environment of each site is important for the discussion of
the magnetic properties of these compounds. We have used the
lattice constants determined in Ref.\,\onlinecite{GalanakisSGS}
using the full-potential nonorthogonal local-orbital minimum-basis
band structure scheme (FPLO)\cite{koepernik} with the GGA
exchange-correlation potential.\cite{GGA} For Mn$_2$CoAl the
calculated value of 5.73 \AA\ is slightly smaller than the
experimentally determined lattice constants of 5.8388 and 5.798
\AA\ in Refs.\,\onlinecite{Liu08} and \onlinecite{Ouardi},
respectively. We compute the ground state electronic properties
using the  full-potential linearized augmented plane wave (FLAPW)
method as implemented in the \texttt{FLEUR} code \cite{Fleur}
combined with the generalized gradient approximation (GGA) to the
exchange-correlation potential as parameterized by Perdew
\textit{et al}.\cite{GGA}

\subsection{Exchange constants and spin-wave dispersion}
\label{sec2_3}

Next we present the formalism on which the calculations of the
exchange constants are based. Our starting point is the classical
Heisenberg model with unit vectors ${\bf e}_{n\alpha}$, pointing
along the spin moments at positions ${\bf R}_{n\alpha}$ specified
by a cell index $n$ and magnetic lattice index $\alpha$. The spin
moments interact via exchange coupling parameters
$J^{\alpha\beta}_{mn}$  and the exchange Hamiltonian $H_{ex}$ is
\begin{equation}
H_{ex}=-\frac{1}{2}\sum_{mn\alpha\beta}J^{\alpha\beta}_{mn}{\bf
  e}_{m\alpha}\cdot {\bf e}_{n\beta}
\label{eq:heisenberg}
\end{equation}
Note that the magnitude of the moments is incorporated in the
parameters $J$. The exchange parameters are extracted by a least
square fit from ab-initio total energy calculations performed
using the FLEUR code for a set of spin spirals with randomized
wave-vectors.\cite{Fleur} To reduce the computational cost we do
non-self consistent calculations and apply the approximation based
on the magnetic force theorem\cite{force-theorem,force-theorem2}
to obtain
 the total energy differences from the differences in sums of
 eigenvalues.\

In order to obtain the adiabatic magnon
dispersion,\cite{magnon1,magnon2,magnon3,magnon4} we set up the
spin-wave dynamical matrix $\Delta({ \bf q})$ at a Brillouin zone
point $\mathbf{q}$ and solve for eigenvalues which provides the
magnon frequencies:
\begin{equation}
\Delta_{\alpha \beta} ({\bf q}) = 2\left ( \delta_{\alpha \beta} \sum_{\gamma}
\frac{ J^{\alpha\gamma}({\bf 0 }) M_{\gamma}}{| M_{\gamma}|| M_{\alpha}|}-\frac{J^{\alpha\gamma}({\bf q })
M_{\beta}}{| M_{\beta}|| M_{\alpha}|}\right)\\
\end{equation}
\begin{equation}
J^{\alpha \beta}({\bf q})=\sum_{n} J^{\alpha\beta}_{0n}\cos[{\bf q} \cdot ( {\bf R}_{0\alpha}- {\bf R}_{n\beta})]
\end{equation}
Here $M_{\alpha}$ is the integrated magnetic moment of sub-lattice
$\alpha$. In order to obtain the spin-stiffness $D$ of the
compounds we fit a linear or quadratic form $D|{\bf q}|$ or
$D|{\bf q}|^{2}$ respectively to the adiabatic magnon energies
along a high symmetry line in the neighborhood of the ${\bf
\Gamma}$-point (in cubic systems, $D$ is isotropic). A linear
behavior is present for the conventional antiferromagnets and we
find it here for the compensated ferrimagnets. We should also note
that within our formalism, where we consider an adiabatic approach
for the magnons, we cannot study the Landau damping of the
spin-waves induced by electron-hole excitations. In the compounds
under study we expect the adiabatic approximation to provide
reasonable results since there is a Stoner gap separating the
magnon spectra from the continuum Stoner excitation spectra as in
most half-metallic magnets (with the exception when the Fermi
level is exactly at the higher energy edge of the minority-spin
energy gap).

\subsection{Temperature dependence of the magnetization and $T_{\mathrm{c}}$}
\label{sec2_4}

We employ the classical Monte Carlo technique to calculate the
temperature dependence of the magnetization that is derived from
the Heisenberg exchange Hamiltonian (\ref{eq:heisenberg}). The
technique provides an excellent estimation of the critical
temperature, $T_\mathrm{c}^\mathrm{MC}$, from the position of the
peak of the susceptibility as a function of temperature calculated
as $\chi(T)=[\langle M^2(T)\rangle-\langle M(T)\rangle^2]/k_BT$,
where $\langle\cdots\rangle$ denotes thermal averaging over Monte
Carlo steps, $M(T)$ is the magnetization (before averaging), and
$k_B$ is the Boltzmann constant. Use of a correction for
finite-size effects, e.g. the fourth-order cumulant
method,\cite{LandauBinder} can give a more accurate estimation of
the critical temperature, but the correction here is small (of the
order of a few K) since we are using large simulation supercells
($12\times12\times12$ primitive cells corresponding to 5184
magnetic atoms). We employ the Metropolis
algorithm\cite{Metropolis} and as a random number generator we use
the Mersene Twister.\cite{Mersene} The input comprises both the
inter-sublattice and intra-sublattice exchange constants
considering only the magnetic atoms, neglecting any contribution
due to the interaction of the low-moment sp atoms. Moreover we
ignore the contribution of the interstitial
region.\cite{interstitial}

From the exchange constants the Curie critical temperature
$T_\mathrm{c}$ can be also estimated within the mean-field
approximation (MFA). Actually, the critical temperature is given
by the average value of the magnon energies which in MFA is the
arithmetic average taking all the magnon values with equal weight.
Thus is an arithmetic property that the MFA estimation is larger
than experimental values \cite{SasiogluH1,pajda,SasiogluH2}. The
MFA expression of the critical temperature for a multi--sublattice
material like Heusler compounds has been provided in literature.
\cite{Sasioglu2004,Anderson}.

\section{Results and Discussion}
\label{sec3}

This section is divided into three parts. In the first part we
discuss the  ground state properties and magnetic moments  of the
studied compounds. The second part deals with the exchange
interactions, spin-wave dispersions, as well as  spin stiffness
constants. The last part  focusses on the temperature dependence
of the magnetization and critical temperature. Note that
Ti$_2$CoSi presents similar behavior to Mn$_2$CoAl and Ti$_2$VAs
presents similar properties to Ti$_2$MnAl. Thus we will focus our
discussion mainly on Mn$_2$CoAl, Ti$_2$MnAl, and Cr$_2$ZnSi
compounds.

\begin{figure*}[t]
\includegraphics[width=\textwidth]{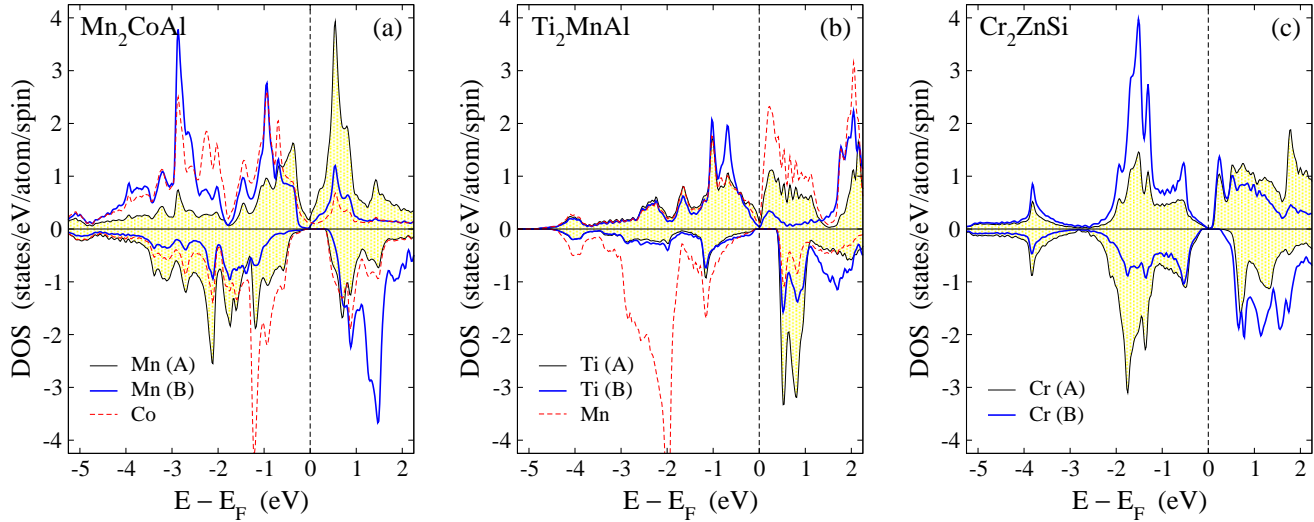}
\vskip -0.3cm \caption{Density of states (DOS) projected on the
transition metal atoms. In the case of Mn$_2$CoAl, positive
(negative) DOS values correspond to the majority (minority) spin
electrons. In the case of the other two compounds, which are
fully-compensated ferrimagnets, spin-up (positive DOS) and
spin-down (negative DOS) electrons have been chosen such that the
atomic spin magnetic moments in Table\,\ref{table1} have the right
sign. Fermi level is set to the zero energy value.}\label{DOS}
\end{figure*}

\subsection{Spin  gapless semiconducting behavior and magnetic moments}
\label{sec3_1}

The first step in the study of these materials is to establish
their ground state properties at 0 K as obtained from our
first-principles calculations. In Fig. \ref{DOS} we have plotted
the density of states (DOS) projected on the transition metal
atoms for Mn$_2$CoAl, Ti$_2$MnAl and Cr$_2$ZnSi and in Table
\ref{table1} we present the atomic and total spin magnetic moments
for all five compounds. We do not present band structure plots
since they are known from previous works.\cite{Liu08,GalanakisSGS}
In all studied compounds we get a finite gap in the minority-spin
band structure (negative values of DOS) and a zero-width gap in
the majority-spin band structure. In the case of Mn$_2$CoAl and
Cr$_2$ZnSi the Fermi level is located at the middle of the
minority-spin energy gap and for Ti$_2$MnAl the Fermi level is
located at the left edge of the gap. The position of the Fermi
level within the gap is important with respect to the coupling
between collective and single electron excitations discussed
later. For all five compounds under study the total DOS per
formula unit (f.u.) is similar to the ones calculated in
Ref.\,\onlinecite{GalanakisSGS} with a different full-potential
method and thus spin-gapless semiconducting behavior of these
compounds is a robust prediction of ab-initio electronic structure
calculations.

\begin{table}
\caption{Calculated atom-resolved and total spin magnetic moments
(in $\mu_\mathrm{B}$) for the five spin-gapless semiconducting
inverse Heusler compounds under study having the chemical formula
X$_2$YZ. The superscripts A and B distinguish the two inequivalent
X atoms; we present the sum of the Z spin moment and the
interstitial spin magnetic moments; $abs$ stands for the sum of
the absolute atomic spin magnetic moments.  Note that we have used
the equilibrium lattice constants as calculated in Ref.
\onlinecite{GalanakisSGS}. }
\begin{ruledtabular}
\begin{tabular}{lrrrrrrr}
Compound &    a(\AA) & m$_{[X^\mathrm{A}]}$    &
m$_{[X^\mathrm{B}]}$   & m$_{[Y]}$ & m$_{[Z+inter]}$ & m$_{[tot]}$  & m$_{[abs]}$   \\
\hline
Mn$_{2}$CoAl  & 5.73  &  $-$1.52   &  2.61   &  0.98 &  0.18& 2.0  & 5.29 \\
Ti$_{2}$CoSi    & 6.03  &   1.41  &  0.71 &  0.39  & 0.49&  3.0 & 3.00\\
Ti$_{2}$MnAl   & 6.24  &   1.13  &  1.00  & $-$2.59  & 0.46& 0.0  & 5.18 \\
Ti$_{2}$VAs     & 6.23  &   1.04  &  0.42 & $-$1.61  & 0.15& 0.0
&3.22 \\ Cr$_{2}$ZnSi   & 5.85  &  $-$1.59   & 1.64  & 0.03 &
$-$0.08& 0.0  & 3.34

\end{tabular}
\label{table1}
\end{ruledtabular}
\end{table}

Among the five studied compounds, Mn$_2$CoAl is a ferrimagnet  and
Ti$_2$CoSi a ferromagnet with total spin magnetic moments per f.u.
of 2 and 3 $\mu_\mathrm{B}$, respectively, and the other three
compounds combine the SGS character to a fully-compensated
ferrimagnetic state presenting zero total spin magnetic moments
per f.u. as shown in Table \ref{table1}. In the case of the four
ferrimagnetic compounds, the X atoms at the B sites couple :
either (a)  antiferromagnetically to the X atoms at the A sites
and ferromagnetically to the Y atoms at the C sites (cases of
Mn$_2$CoAl and Cr$_2$ZnSi), or (b) ferromagnetically to the X
atoms at the A sites and antiferromagnetically to the Y atoms at
the C sites (cases of Ti$_2$MnAl and Ti$_2$VAs). This behavior is
expected from the so-called Bethe-Slater curve.\cite{Bethe-Slater}
The early transition metal atoms like, Cr and Mn, when they are
close to each other in space tend to have antiparallel spin
magnetic moments. On the other hand for the nearest Ti atoms, the
coupling tends to be  ferromagnetic (Ti$^A$ and Ti$^B$ atoms in
Ti$_2$YZ compounds). This behavior is also reflected on the
exchange constants calculated and presented in the next
subsection. Interestingly even in the case of Cr$_2$ZnSi, where Zn
is almost non-magnetic since all its $d$-states are occupied lying
below the energy window presented in Fig. \ref{DOS}, the small
induced spin magnetic moment at the Zn and Si atoms leads to a
small imbalance of the spin moments between the two Cr atoms
(Cr$^A$ has as a spin moment of $-$1.59 $\mu_\mathrm{B}$ and
Cr$^B$ of 1.64 $\mu_\mathrm{B}$) and Cr$^A$ resolved DOS presented
in Fig. \ref{DOS} is not an exact mirror image of the Cr$^B$ DOS
as in conventional antiferromagnets.

Finally, we should shortly discuss the values of the atomic spin
magnetic moments presented in Table\,\ref{table1}. For all five
compounds, spin moments are similar to the results in
Ref.~\onlinecite{GalanakisSGS} where the FPLO(\cite{koepernik}
electronic structure code has been employed. Moreover for
Mn$_2$CoAl results agree with the calculated values in Refs.
\onlinecite{Liu08,Meinert11,Ouardi} and for Ti$_2$MnAl with the
results presented in Ref. \onlinecite{Jia2014}. Since in each
study a different full-potential ab-initio method has been used,
we can be confident of the validity of our results. Concerning now
the experimentally available data, in Refs.\,\onlinecite{Liu08}
and \onlinecite{Ouardi} only the total spin magnetic moment per
f.u. for Mn$_2$CoAl has been measured which has been found to be
1.95 and exactly 2 $\mu_\mathrm{B}$, respectively, in agreement
with our calculated value of 2 $\mu_\mathrm{B}$. A similar total
spin magnetic moment (1.94 $\mu_\mathrm{B}$) has been measured by
Xu and collaborators \cite{Xu2014} for Mn$_2$CoAl films on Si
substrates at 5 K. The only discrepancy occurs in the study of
Jamer \textit{et al.}\cite{Jamer14} in Mn$_2$CoAl films on
GaAs(001), where X-ray magnetic circular dichroism (XMCD)
experiments where carried out. Although XMCD is a powerful
technique, in Mn$_2$CoAl there exist two Mn atoms with opposite
spin magnetic moments. XMCD can distinguish between different
elements but cannot distinguish atoms of the same chemical element
with different spin magnetic moments and thus the values of the Mn
moments in Ref. \onlinecite{Jamer14} cannot be interpreted as
atomic moments. Note that in FLAPW method, the atomic spin
magnetic moments are calculated by integrating the spin-dependent
charge density within each muffin-tin sphere surrounding each
atom. The interstitial region is not assigned to any atom.
Although we have used almost touching muffin-tin spheres we can
see in Table\,\ref{table1} that in the case of Ti$_2$MnAl and
Ti$_2$CoSi a significant part of the spin magnetic moment is
located at the interstitial region. This should be attributed to
the fact that early transition metals have $d$-states extending
further away from the nucleus compared to late transition metals
and the smaller the valence the more extended are the $d$-states.

\begin{figure*}[t]
\includegraphics[width=\textwidth]{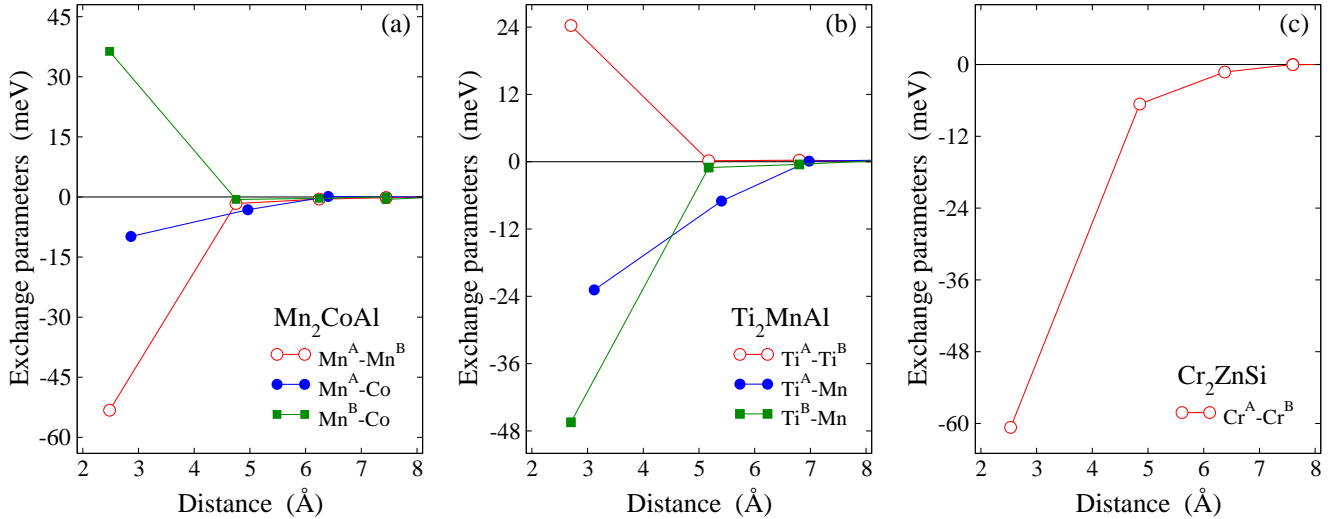} \vskip
-0.2cm \caption{Inter-sublattice Heisenberg exchange parameters as
a function of inter-atomic distances for (a) Mn$_2$CoAl, (b)
Ti$_2$MnAl,and (c) Cr$_2$ZnSi.}\label{ExchConst}
\end{figure*}

\begin{table*}
\caption{On-site intra-sublattice (\textit{e.g.}
$J_{0}^{\textrm{X$^A$-X$^A$}}\equiv\sum_{\mathbf{R}}J_{\mathbf{0R}}^{\textrm{X}^{A}-\textrm{X}^{A}}$,
where $\mathbf{R}$ is the lattice vector) and inter-sublattice
(\textit{e.g.}
$J_{0}^{\textrm{X$^A$-X$^B$}}\equiv\sum_{\mathbf{R}}J_{\mathbf{0R}}^{\textrm{X}^{A}-\textrm{X}^{B}}$)
exchange constants (in meV),  spin-wave stiffness constant $D$ (in
meV$\cdot$\AA\  for compensated ferrimagnets or meV$\cdot$\AA$^2$
for ferri- and ferromagnets), and the mean-field and Monte Carlo
calculated critical temperatures (in K) for the compounds under
study. Note that for the inverse X$_2$YZ compounds the two X
transition metal atoms occupy the A and B sites and the third
transition metal atom Y occupies the C site.}
\begin{ruledtabular}
\begin{tabular}{lrrrrrrrrrrrrrrrrrrrrrr}
Compound   & $J_{0}^{\textrm{X$^A$-X$^A$}}$ &
$J_{0}^{\textrm{X$^B$-X$^B$}}$ & $J_{0}^{\textrm{Y-Y}}$ &
$J_{0}^{\textrm{X$^A$-X$^B$}}$ & $J_{0}^{\textrm{X$^A$-Y}}$ &
$J_{0}^{\textrm{X$^B$-Y}}$ &  $D$$\:\:$ &
$T_{\textrm{c}}^{\textrm{MFA}}$ (K) &
$T_{\textrm{c}}^{\textrm{MC}}$ (K) \\ \hline
Mn$_{2}$CoAl   & $-$73.0  & $-$29.7   & $-$10.5   & $-$238.4 & $-$84.5    &  132.4  & 677 meV$\cdot$\AA$^2$  & 1134 &  770 \\
Ti$_{2}$CoSi   & 22.7   &  6.9    & 5.22    & 169.4  & 69.8     &  0.6    & 636  meV$\cdot$\AA$^2$ &766  &  550 \\
Ti$_{2}$MnAl   &  6.5   &  1.9    &  $-$30.0  & 101.4  & $-$184.4   &  $-$196.8 & 274  meV$\cdot$\AA$\:\:$ & 1222 &  960 \\
Ti$_{2}$VAs    &  37.5  &  2.3    & 92.4    & 45.8   & $-$145.5&
$-$55.1 & 598 meV$\cdot$\AA$\:\:$ & 910  & 800 \\
Cr$_{2}$ZnSi   &  21.2  & $-$18.1   &    $-$    & $-$336.4 &  $-$       &  $-$      & 752 meV$\cdot$\AA$\:\:$ & 1308 &  1040 \\
\end{tabular}
\label{table2}
\end{ruledtabular}
\end{table*}

\subsection{Exchange interactions and spin-wave dispersion}
\label{sec3_2}

Our calculations have shown that in the compounds under study
inter-sublattice exchange interactions play a dominant role  in
formation of the magnetic state and critical temperature. In
Fig.\,\ref{ExchConst} we present the inter-sublattice exchange
constants as a function of distance. Negative values of the
exchange constants reflect an antiferromagnetic coupling of the
corresponding spin moments and positive values a ferromagnetic
coupling. In all compounds the inter-sublattice nearest neighbor
interactions dominate and especially the interaction between the
X$^B$ atom and its  X$^A$ and Y nearest neighbors
(Fig.\,\ref{Structure}). The interactions between next-nearest
neighbors X$^A$ and Y are expectedly weaker. In the case of
Cr$_2$ZnSi the Zn atom is almost non magnetic while in the case of
Mn$_2$CoAl the intra-sublattice exchange constants between the
Mn$^A$-Mn$^A$ atoms have a sizeable value despite their large
distance.  In the case of Ti$_2$MnAl, the Ti$^{A,B}$ have positive
spin moments and the Mn$^C$ has negative spin moment as shown in
Table\,\ref{table1}. The situation is different to Mn$_2$CoAl
where Mn$^A$ site has negative, and Mn$^B$ site and Co$^C$ have
positive moments. Thus the X$^A$-X$^B$ and X$^B$-Y interactions
have different signs for the two compounds. We can conclude that
for the SGSs under study the interactions are short range: as seen
from the Fig.\,\ref{ExchConst} they decay quickly with the
distance. This can be attributed to the existence of the finite
spin-gap as discussed in literature in detail for half metallic
Heusler compounds.\cite{Tc0,Tc1}

The short range nature of exchange interaction in  Mn$_2$CoAl and
similar compounds has been shown by Meinert and collaborators in
Ref.\,\onlinecite{Meinert11}. Moreover, in the case of Mn$_2$CoAl
our results agree very well with the results obtained in Ref.\,\onlinecite{Meinert11},
considering the difference of a pre-factor $1/2$ in the definitions of
the Heisenberg Hamiltonian between Ref.\,\onlinecite{Meinert11} and the present work.
(Also note that in their notations the sequence of the atoms is Co-Mn$^B$-Mn$^C$-Al
and thus our Mn$^A$ atom corresponds to the Mn$^C$ atom in their article).
Further calculations by Wollmann and collaborators on a series of Mn$_2$-based
inverse Heusler compounds have confirmed the short-range nature of the
interactions in these materials.\cite{Wollmann}

The contribution of each type of exchange interaction to the total
exchange field $J_{0}$ (the Weiss field acting on a spin moment)
is given by the sum of the interactions over all possible pairs
or, partially, over all pairs within a sublattice. The calculated
values for the intra- and inter-sublattice $J_{0}$ is presented in
Table\,\ref{table2}. Our results confirm the conclusions drawn in
the previous paragraph. The on-site inter-sublattice $J_{0}$ are
considerably larger than the intra-sublattice $J_{0}$ and only in
the case of Mn$_2$CoAl  and Ti$_2$VAs the
$J_{0}^{\textrm{Mn$^A$-Mn$^A$}}$ and $J_{0}^{\textrm{V-V}}$  makes
a considerable contribution into the total exchange field with
different signs.

\begin{figure}[t]
\includegraphics[width=\columnwidth]{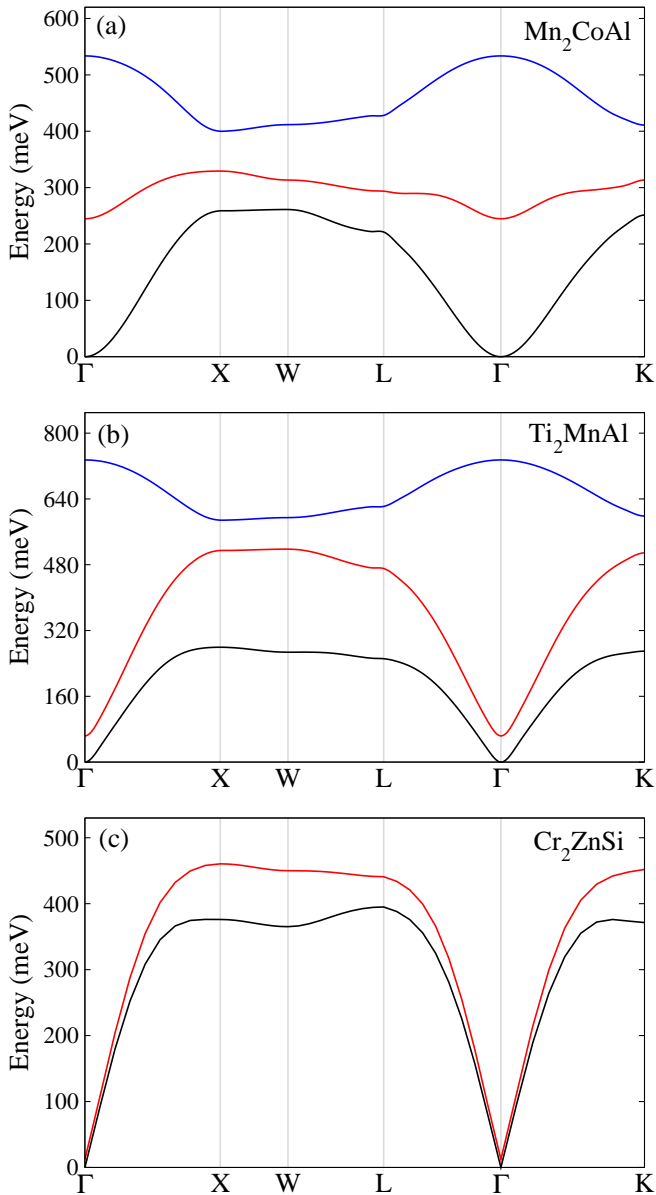}
\vskip -0.1cm \caption{Spin-wave dispersion curves along the high
symmetry lines in Brillouin zone for (a) Mn$_2$CoAl, (b)
Ti$_2$MnAl,and (c) Cr$_2$ZnSi. In the case of Mn$_2$CoAl and
Ti$_2$MnAl, there are three branches since there are three
magnetic atoms per unit cell; the sp-atom are almost non-magnetic.
In the case of Cr$_2$ZnSi, the Zn atoms have all their $d$-valence
states completely occupied and have vanishing spin magnetic
moments, and thus there are two branches. In the case of
Mn$_2$CoAl the dispersion curve of the acoustic branch around the
$\Gamma$ point shows a quadratic behavior while for  Cr$_2$ZnSi
the behavior is linear In the case of Ti$_2$MnAl, it is in
between.}\label{Magnons}
\end{figure}

In Fig.\,\ref{Magnons} we present the spin-wave dispersion  along
the high symmetry lines in the Brillouin zone for Mn$_2$CoAl,
Ti$_2$MnAl and Cr$_2$ZnSi. Each spectrum has distinct features.
First, the number of branches coincides with the number of
magnetic atoms in the unit cell and thus we have three branches
for Mn$_2$CoAl and TI$_2$MnAl and two branches for Cr$_2$ZnSi
where the Zn atom is not magnetic.  The energy dispersion curves
of all compounds under study are typical for magnets with short
range interactions, where nearest-neighbors and
next-nearest-neighbors interactions dominate, and do not yield any
instabilities. Instabilities can occur if the acoustic magnon
modes have very low energies (close to zero)  in some parts of the
Brillouin zone but this is not the case for any of the studied
compounds.

In the case of Mn$_2$CoAl the acoustic branch shows a typical
behavior of ferro/ferrimagnets and around the $\Gamma$ point the
energy-dispersion curve shows a quadratic behavior with a
spin-wave stiffness constant $D$ of 677 meV$\cdot$\AA$^2$. This
value exceeds typical values of transition metal ferromagnets
which usually range between 300 and 600
meV$\cdot$\AA$^2$,\cite{pajda} and is close to the maximum known
values of 715 meV$\cdot$\AA$^2$ for Co$_2$FeSi\cite{D1} and 800
meV$\cdot$\AA$^2$ for Fe$_{53}$Co$_{47}$.\cite{D2} Ti$_2$CoSi as
shown in Table \ref{table2} exhibits a $D$ value of 636
meV$\cdot$\AA$^2$ close to the value for Mn$_2$CoAl.

Although Cr$_2$ZnSi is not a true antiferromagnet, around the
$\Gamma$ point the energy dispersion is linear and the optical and
acoustic branch coincide showing considerable difference only at
the plateau close to the Brillouin zone boundary (X, W, L, K
points in Fig.~\ref{Magnons}). The spin-wave stiffness constant as
shown in Table\,\ref{table2} takes a value of 752 meV$\cdot$\AA.
 In the case of Ti$_2$MnAl the
situation is more complicated. Due to the small spin moments of
the Ti atoms around the $\Gamma$ point the energy-dispersion of
the acoustic magnon is practically linear but with a small value
of the $D$ constant (274 meV$\cdot$\AA). Ti$_2$VAs also shows a
behavior similar to Ti$_2$MnAl but now $D$ has a much higher value
of 598 meV$\cdot$\AA.

\subsection{Temperature dependence of magnetization and $T_{\rm c}$
\label{sec3_3}}

\begin{figure*}[t]
\includegraphics[width=\textwidth]{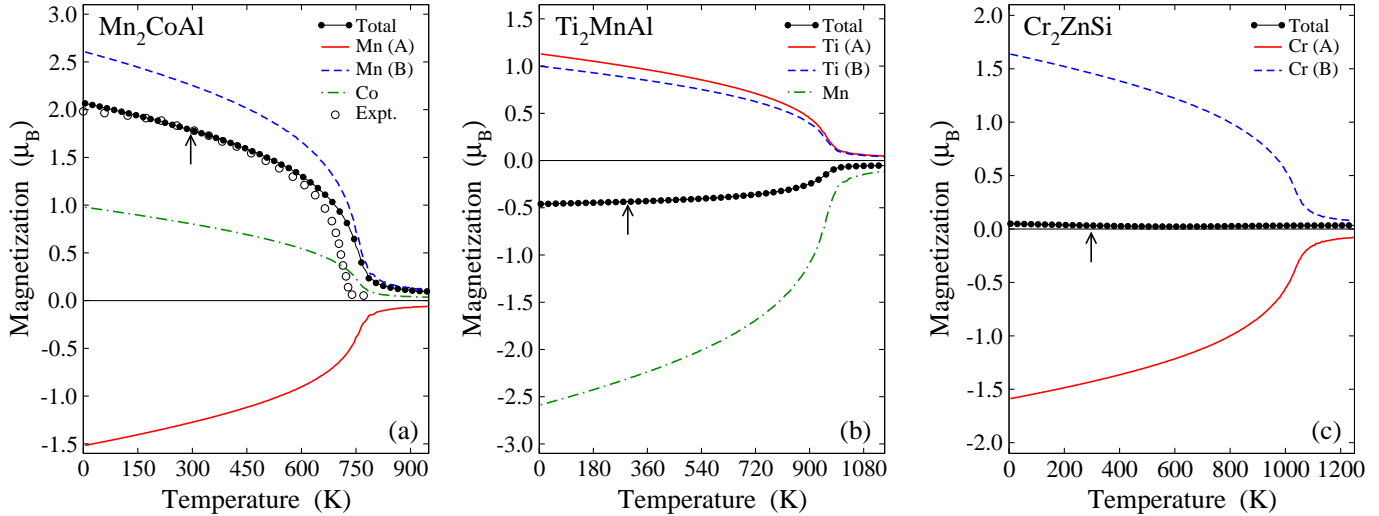}
\vskip -0.2cm \caption{ Calculated temperature dependence of the
sublattice and total magnetization for (a) Mn$_2$CoAl, (b)
Ti$_2$MnAl, and (c) Cr$_2$ZnSi. In the case of Ti$_2$MnAl the
total spin magnetic moment for 0 K is $-$0.46 $\mu_\mathrm{B}$
since in the Monte-Carlo calculations we ignore the sp-atoms as
well as the interstitial region (see Table\,\ref{table1}). With
the arrow we denote the room temperature of 300 K. In the case of
Mn$_2$CoAl we include also the experimental results from
Ref.\,\onlinecite{Ouardi}.} \label{MvsT}
\end{figure*}

In this section we show the temperature dependence of the
magnetization and susceptibility calculated with the classical
Monte-Carlo  technique. In Fig. \ref{MvsT} we present for
Mn$_2$CoAl, Ti$_2$MnAl and Cr$_2$ZnSi the temperature dependence
of the sublattice and total magnetization per f.u. First, we
compare our results to experiment. We plot for Mn$_2$CoAl our
theoretical results together with the experimental results from
the Ref.\,\onlinecite{Ouardi} where the temperature dependence of
the total magnetization in a pollycrystalline film was measured in
a field of 1 T. The agreement between the two data sets is good
with one curve falling on top of the other with the exception of
the region close to the critical temperature. Our curve shows an
abrupt decrease close to our calculated critical temperature of
770 K, while in experiments this sharp decrease is shifted lower
in temperature since the measured critical temperature in Ref.
\onlinecite{Ouardi} is 720 K, slightly smaller than our value.
Thus the Monte-Carlo technique, which we employ, accurately
describes the temperature dependence of the magnetization,
indicating also the accurate determination of the exchange
constants.

In the case of Cr$_2$ZnSi, the picture is that of an
antiferromagnet with a total spin magnetic moment being almost
equal to zero for all temperatures, since we have ignored the
magnetic properties of the Zn atom. In the case of Ti$_2$MnAl the
calculated total magnetization (Fig. \ref{MvsT}) is not exactly
zero as expected from the first-principles result at 0 K
(Table\,\ref{table1}) but it equals to $-$0.46 $\mu_B$ due to the
spin moment of the Al atom and of the interstitial region which we
have ignored in our Monte-Carlo calculations. At and close to room
temperature (up to $\approx 400$~K), where most devices operate,
we find that all three compounds still have sizeable values of the
sublattice magnetization. Thus the magnetic properties are not
deteriorated although the spin-gapless behavior could be lost at
this elevated temperature. For example for Mn$_2$CoAl at 0 K the
Mn$^A$, Mn$^B$ and Co atoms show a sublattice magnetization of
about $-1.5$, 2.6 and 1.0 $\mu_\mathrm{B}$ respectively. At room
temperature, these values become $-1.3$, 2.3 and 0.8
$\mu_\mathrm{B}$, showing an absolute-value decrease of 13\% ,
11\% and 18\% , respectively. The total magnetization per f.u.
decreases from 2.0 $\mu_\mathrm{B}$ at 0 K to 1.8 $\mu_\mathrm{B}$
at room temperature showing an even smaller decrease of 10 \%.
Thus the compounds under study are adequate to be employed in
spintronic/magnetoelectronic devices.

\begin{figure}[t]
\includegraphics[width=\columnwidth]{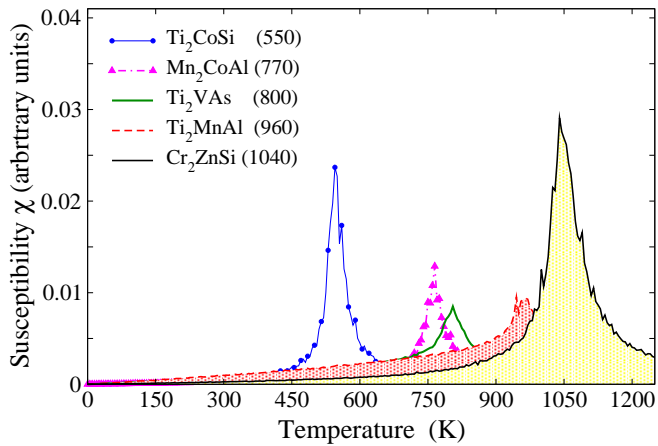}
\vskip -0.2cm \caption{ Calculated temperature dependence of the
susceptibility for SGSs. The maximum of the susceptibility
corresponds to the critical temperature (shown in parenthesis in
the legends).}\label{XvsT}
\end{figure}

To estimate the critical temperature $T_\mathrm{c}^\mathrm{MC}$
within the Monte-Carlo technique  we have plotted in
Fig.\,\ref{XvsT} the susceptibility $\chi$ versus the temperature
for all compounds under study.  Obtained values are  presented in
Table\,\ref{table2} and compared with the  mean-field estimation
of the critical temperature $T_\mathrm{c}^\mathrm{MFA}$.  As seen
for all compounds $T_\mathrm{c}^\mathrm{MC}$ is much lower than
the $T_\mathrm{c}^\mathrm{MFA}$. The difference ranges from 110 K
for Ti$_2$VAs up to 364 K for Mn$_2$CoAl, while for the other
compounds it is around 220-270 K. Note that within MFA, the
critical temperature is given by the arithmetic average of all the
magnon energy values with equal weight. In reality low-energy
magnons have more weight in determining the critical temperature
and MFA usually overestimates experimental data by more than
20\%.\cite{SasiogluH1,pajda,SasiogluH2} The Monte-Carlo determined
critical temperature does not suffer from this drawback and thus
it is expected to approach the experimental values of the critical
temperature.

In some magnetic compounds, an empirical trend is observed that the
critical temperature increases with the magnitude of the local
magnetic moments.\cite{Tc1,Tc2,Tc3,Tc4} This trend is not observed
here. The sum of absolute local moments (see Table \ref{table1}) is
largest for Mn$_2$CoAl being 5.29 $\mu_\mathrm{B}$ but the critical
temperature is largest for Cr$_2$ZnSi for which this sum is only 3.34
$\mu_\mathrm{B}$. Intuitively we can explain this observation as
follows. The critical temperature depends both (a) on the nature
of the spin-dependent wavefunction overlap and chemical bonds between
nearest neighbors, which is a short-range property and is independent
of the magnitude of the spin moments, as well as (b) the more distant
interactions which are formed largely because of susceptibility
effects (one atom polarizes the Fermi surface electrons and this is
felt by a distant atom through the electron propagation) which largely
depend on the magnitude of the spin magnetic moments. Here, since the
nearest-neighbor interactions are dominant and the long-range ones
practically vanish, (a) applies but not (b).

Finally, we compare our values for $T_\mathrm{c}$ with
experimental and theoretical results on Mn$_2$CoAl compound. Our
calculated MFA and Monte-Carlo values are 1134 K and 770 K,
respectively. Meinert and collaborators \cite{Meinert11} estimated theoretically a
$T_\mathrm{c}^{\rm MFA}=890$~K,  smaller than our MFA
result. Wollman and collaborators\cite{Wollmann}
calculated $T_\mathrm{c}^{\rm MFA}=985$~K, in-between our
value and the value in Ref.\,\onlinecite{Meinert11} and within the more
accurate spherical approximation (SPA)\cite{SPA} a value of 740 K, close
to our Monte-Carlo value of 770 K.\cite{Wollmann} Experimentally, for
the bulk-like pollycrystalline film, Ouardi and collaborators
measure for the $T_\mathrm{c}$  a value of 720 K,\cite{Ouardi}
while the experiments by Xu and collaborators on thin films of
Mn$_2$CoAl gave a $T_\mathrm{c}$  value of about 550
K;\cite{Xu2014} this discrepancy is expected since films present
critical temperatures significantly smaller that the bulk samples.
Thus our Monte-Carlo value of 770 K is in good agreement both
with the SPA calculated value of 740 K of Wollmann \textit{et
al.}\cite{Wollmann} and with the experimental value of 720 K measured
by Ouardi \textit{et al.}\cite{Ouardi}

\section{Conclusions}
\label{sec4}

We employed first principles electronic structure calculations in
conjunction with the frozen-magnon method to calculate exchange
interactions, spin-wave dispersion, and spin-wave stiffness
constants in inverse-Heusler-based spin gapless semiconductor
(SGS) compounds Mn$_2$CoAl, Ti$_2$MnAl, Cr$_2$ZnSi, Ti$_2$CoSi and
Ti$_2$VAs. We find that the magnetic behavior of the SGSs is
similar to the half-metallic ferromagnetic full-Heusler alloys,
i.e., the inter-sublattice exchange interactions play an essential
role in the formation of the magnetic ground state and in
determining the critical temperature, $T_\mathrm{c}$. All
compounds, except Ti$_2$CoSi, possess a ferrimagnetic ground
state. Due to the finite energy gap in one spin channel, the
exchange interactions decay rapidly, and hence magnetism of these
SGSs can be described considering only nearest and next-nearest
neighbor exchange interactions. The calculated spin-wave
dispersion curves are typical for ferrimagnets and ferromagnets,
i.e., the number of inequivalent spin-wave branches in the
dispersion curves is equal to the number of magnetic atoms in the
unit cell. Due to the short range nature of the exchange
interactions, the calculated spin-wave stiffness constants turn
out to be larger than the elementary 3$d$-ferromagnets.

Calculated exchange parameters are used as input to determine the
temperature dependence of the magnetization and the critical
temperature $T_\mathrm{c}$ of the SGSs. We find that the
$T_\mathrm{c}$ of  all compounds is much above the room
temperature. The calculated magnetization curve for Mn$_2$CoAl as
well as the critical temperature are in good agreement with
available experimental data.

Our results suggest that, except Mn$_2$CoAl which has been already
synthesized, there are other potential SGS presenting also very
high Curie temperatures. In SGS materials, only one spin channel
contributes to the transport properties, whereas the other spin
channel allows for tunable charge-carrier concentrations. Among
the five studied SGS compounds Ti$_2$MnAl, Ti$_2$VSi and
Cr$_2$ZnSi present also zero magnetization thus creating vanishing
stray fields   leading to advantages that have been pointed out
for antiferromagnetic spintronic devices.\cite{AFM1,AFM2} Thus we
expect such fully-compensated ferrimagnetic compounds to be more
adequate for room temperature spintronic/magnetoelectronic
applications based on spin transport.

\section*{Acknowledgements}
This work was partly supported by the Young Investigators Group
Programme of the Helmholtz Association, Germany, contract
VH-NG-409. We gratefully acknowledge the support of J\"ulich
Supercomputing Centre (grant jiff38).

\end{document}